\documentclass[pra,twocolumn,showpacs]{revtex4}
\usepackage{amsmath,amssymb}
\usepackage{graphicx}
\usepackage{graphics}
\usepackage{bm}
\usepackage{color}
\usepackage{verbatim,color,ulem}

\begin{document}

\title{Localized modes in dense repulsive and attractive Bose-Einstein
condensates \\ with spin-orbit and Rabi couplings}
\author{Luca Salasnich$^{1}$ and Boris A. Malomed$^{2}$}
\affiliation{$^1$Dipartimento di Fisica e Astronomia ``Galileo Galilei'' 
and CNISM, Universit\`a di Padova, Via Marzolo 8, 35131 Padova, Italy \\
$^{2}$Department of Physical Electronics, School of Electrical Engineering,
Faculty of Engineering, Tel Aviv University, Tel Aviv 69978, Israel}

\begin{abstract}
We consider a binary Bose-Einstein condensate with linear and nonlinear
interactions between its components, which emulate the spinor system with
spin-orbit (SO) and Rabi couplings. For a relatively dense condensate, 1D
coupled equations with the nonpolynomial nonlinearity of both repulsive and
attractive signs are derived from the 3D Gross-Pitaevskii equations.
Profiles of modes confined in an external potential under the action of the
self-repulsion, and self-trapped solitons in the case of the
self-attraction, are found in a numerical form and by means of analytical
approximations. In the former case, the interplay of the SO and Rabi
couplings with the repulsive nonlinearity strongly distorts shapes of the
trapped modes, adding conspicuous side lobes to them. In the case of the
attractive nonlinearity, the most essential result is reduction of the
collapse threshold under the action of the SO and Rabi couplings.
\end{abstract}

\maketitle


\section{Introduction}

The recently proposed emulation of the spin-orbit (SO) coupling in condensed
matter by means of similar but weak (i.e., theoretically tractable and
experimentally controllable) interactions in binary Bose-Einstein
condensates (BECs) \cite{SO-BEC} has drawn much attention, in the spirit of
using dilute Bose gases as \textit{quantum simulators} \cite{simulator}. In
this context, BEC composed as mixtures of different atomic states may
represent \textit{pseudo-spin} systems, with the multi-component mean-field
wave functions emulating the spinor order parameter. In experiments, the
artificial SO coupling has been implemented in both bosonic \cite%
{so-bose,so-bose2} and fermionic \cite{so-fermi1,so-fermi2} atomic gases, by
means of counterpropagating laser beams which couple two internal hyperfine
states of the atom by a stimulated two-photon Raman transition. In this
connection, it is relevant to mention that the true SO coupling (rather than
the emulated one) is possible too in BEC composed of spinor bosons \cite%
{true-SO} and in magnon condensates \cite{magnon-SO}.

The single-particle SO Hamiltonian which can be implemented in the BEC is
\begin{equation}
{\hat{h}}_{\mathrm{sp}}=\left[ {\frac{{\hat{\mathbf{p}}}^{2}}{2m}}+U(\mathbf{%
r})\right] +{\frac{\hbar \Omega }{2}}\sigma _{x}-{\frac{k_{L}}{m}}\ {\hat{p}}%
_{x}\sigma _{z},  \label{hsp}
\end{equation}%
where ${\hat{\mathbf{p}}}=-i\hbar (\partial _{x},\partial _{y},\partial _{z})
$ is the momentum operator, $U(\mathbf{r})$ is a trapping potential, $k_{L}$
is the recoil wavenumber induced by the interaction with the laser beams, $%
\Omega $ is the frequency of the Raman coupling, which is responsible for
the Rabi mixing between the two states, and $\sigma _{x,z}$ are the Pauli
matrices. Recently, considerable attention has been drawn to models
combining the SO coupling, which is a linear feature, and mean-field
nonlinearities, which are induced, as usual \cite{TF}, by inter-atomic
collisions. Various dynamical effects have been investigated in the
framework of such nonlinear systems, including the self-trapping and
formation of solitons \cite{so-bright1,so-bright2}, vortical patterns \cite%
{SO-vortices}, the interplay of the SO coupling and dipole-dipole
interactions \cite{SO-dipolar}, the modulational instability of uniform
states in SO-coupled BEC \cite{MI}, etc.

The objective of the present work is to present an effective 1D model for
the SO- and Rabi-coupled \textit{relatively dense} binary BEC, and
investigate basic properties if trapped modes in the framework of the model.
The \textit{dimension reduction}, i.e., derivation of the 1D system from the
underlying set of the 3D Gross-Pitaevskii equations (GPEs) with cubic terms,
leads to nonpolynomial nonlinearity, as was previously demonstrated in the
context of single- \cite{NPSE} and binary \cite{sala-boris} condensates with
the attractive nonlinearity, as well as for the repulsive nonlinearity \cite%
{Delgado}. In the former case, the corresponding nonpolynomial
nonlinear Schr\"{o}dinger equations (NPSEs)\ generate solitons, as
the usual 1D GPEs with the postulated cubic nonlinearity. However,
on the contrary to the 1D cubic equations, the NPSEs predict the
\emph{onset of collapse} in the effectively 1D solitons at a
critical strength of the self-attraction \cite{NPSE}. This feature
makes the 1D solitons similar to 3D solitons, which are also subject
to the collapse above a critical value of the number of atoms,
depending on parameters of the BEC-trapping potential. Detailed
numerical investigations have demonstrated that the near-collapse
dynamics of 3D solitons (although not the collapse itself) is quite
accurately approximated by their 1D counterparts in the framework of
the NPSE \cite{NJP}.

In this work, we derive a system of two NPSEs coupled by linear and
nonlinear terms, which include new terms accounting for the emulated SO
interaction. The derivation is performed for both the repulsive and
attractive signs of the inter-atomic interactions, and is followed by the
analysis of trapped modes, for both signs of the nonlinearity. In the case
of the self-repulsion, the trapping is imposed by an axial
harmonic-oscillator (HO) potential, while the attractive condensates
self-traps into solitons. The analysis concentrates on new features of the
externally trapped and self-trapped modes in the presence of the
SO-emulating interactions.

The rest of the paper is organized in the following way. The derivation of
the NPSE\ system for the binary condensate with the SO and Rabi interactions
between the components is reported in Section II. This section also includes
analytical approximations which help to understand characteristic features
of the trapped modes in the SO-coupled dense condensates. Basic numerical
results are reported in Section III. In both cases of the self-repulsion and
self-attraction, the nonpolynomial nonlinearity strongly affects shapes of
the trapped modes and solitons. The most essential result is a \textit{%
reduction} of the collapse threshold for the solitons under the action of
the SO and Rabi interactions. The paper is concluded by Section IV.

\section{Spin-orbit coupled nonpolynomial Schr\"{o}dinger equations}

\subsection{Derivation of the model}

Our starting point is the 3D version of the SO system which was recently
introduced in the form of 1D GPEs with the cubic nonlinearity \cite%
{so-bright1}. In fact, the 1D equations postulated in Ref.
\cite{so-bright1} are essentially the same as those derived earlier
in the context of nonlinear fiber optics, for two polarizations of
light co-propagating in a twisted birefringent fiber \cite{old}). A
different version of the 1D SO Hamiltonian was adopted in another
recent paper \cite{so-bright2}. If one takes into account the
nonlinearity induced by atomic collisions, the present model and the
one studied in Ref. \cite{so-bright2} can be transformed into each
other, provided that the strength of the nonlinear terms is
characterized by the single \textit{s}-wave scattering length.

The SO- and Rabi-coupled binary BEC, confined in the plane of $(y,z)$ by a
tight HO potential with trapping frequency $\omega _{\bot }$, and in the $x$
direction by a generic loose potential $V(x)$, is modeled by the system of
3D GPEs for macroscopic wave functions $\psi _{k}$ of the two atomic states (%
$k=1,2$):
\begin{gather}
i\,\partial _{t}\psi _{k}=\Big[-{\frac{1}{2}}\nabla ^{2}+V(x)+{\frac{1}{2}}%
(y^{2}+z^{2})+(-1)^{k-1}i\gamma \partial _{x}  \notag \\
+2\pi \,g_{k}|\psi _{k}|^{2}+2\pi \,g_{12}|\psi _{3-k}|^{2}\Big]\psi
_{k}+\Gamma \ \psi _{3-k},  \label{GPE}
\end{gather}%
where the lengths, time, and energy are measured in units of $a_{\bot }=%
\sqrt{\hbar /(m\omega _{\bot })}$, $\omega _{\bot }^{-1}$, and $\hbar \omega
_{\bot }$, respectively. Here $g_{k}\equiv 2a_{k}/a_{\bot }$, $g_{12}\equiv
2a_{12}/a_{\bot }$ are strengths of the intra- and inter-species
interactions, with the respective scattering lengths $a_{k}$ and $a_{12}$,
while $\gamma \equiv k_{L}a_{\bot }$ and $\Gamma \equiv \Omega /(2\omega
_{\bot })$ are dimensionless strengths of the SO and Rabi couplings,
respectively [recall that $k_{L}$ is the recoil wavenumber in Hamiltonian (%
\ref{hsp})]. The time-dependent number of atoms in the $k$-th state is $%
N_{k}(t)=\int \int \int \ dxdydz\ |\psi _{k}(x,y,z,t)|^{2}$, the constant
total number of atoms being $N=N_{1}(t)+N_{2}(t)$.

To reduce the dimension from 3D to 1D, we adopt the usual factorized ansatz
for the wave functions which are tightly trapped in the transverse plane, $%
(y,z)$, and weakly confined in the axial direction, $x$ \cite%
{NPSE,sala-boris}:
\begin{equation}
\psi _{k}(x,y,z,t)={\frac{1}{\sqrt{\pi }\eta _{k}(x,t)}}\exp {\left\{ -{%
\frac{y^{2}+z^{2}}{2\eta _{k}^{2}(x,t)}}\right\} }\,f_{k}(x,t)\;,
\label{ansatz}
\end{equation}%
where $\eta _{k}(x,t)$ and $f_{k}(x,t)$ are the transverse widths and axial
wave functions, respectively, the latter normalized by conditions
\begin{equation}
\int_{-\infty }^{+\infty }\left\vert f_{k}(x,t)\right\vert ^{2}dx=N_{k}(t),
\label{N}
\end{equation}%
which are compatible with Eq. (\ref{ansatz}). The conserved total number of
atoms is $N=N_{1}+N_{2}$.

Inserting ansatz (\ref{ansatz}) into the Lagrangian density which produces
Eqs. (\ref{GPE}), performing the integration in the transverse plane, and
neglecting, as usual, derivatives of $\eta _{k}(x,t)$, one can derive the
corresponding effective Lagrangian, which then gives rise to a system of
four variational equations:
\begin{gather}
i\,\partial _{t}f_{k}=\Big[-{\frac{1}{2}}\partial
_{x}^{2}+V(x)+(-1)^{k-1}i\gamma \partial _{x}+{\frac{1}{2}}({\frac{1}{\eta
_{k}^{2}}}+\eta _{k}^{2})+  \notag \\
{\frac{g_{k}}{\eta _{k}^{2}}}|f_{k}|^{2}+2{\frac{g_{12}}{(\eta _{1}^{2}+\eta
_{2}^{2})}}|f_{3-k}|^{2}\Big]f_{k}+2\Gamma {\frac{\eta _{1}\eta _{2}}{(\eta
_{1}^{2}+\eta _{2}^{2})}}f_{3-k}\;,  \label{f}
\end{gather}%
\begin{gather}
\eta _{k}^{4}=1+{g_{k}|f_{k}|^{2}}+{4g_{12}|f_{3-k}|^{2}}{\frac{\eta _{k}^{4}%
}{(\eta _{1}^{2}+\eta _{2}^{2})^{2}}}  \notag \\
+2(-1)^{k-1}\Gamma {\frac{(f_{1}^{\ast }f_{2}+f_{2}^{\ast }f_{1})}{%
|f_{1}|^{2}}}{\frac{\eta _{k}^{3}\eta _{3-k}(\eta _{1}^{2}-\eta _{2}^{2})}{%
(\eta _{1}^{2}+\eta _{2}^{2})^{2}}}.  \label{eta}
\end{gather}%
In most cases, a reasonable assumption is that strengths of the nonlinear
interactions between different species are equal,
\begin{equation}
g_{1}=g_{2}=g_{12}\equiv g  \label{ggg}
\end{equation}%
\cite{gg} [in fact, a more general case is considered below too, see Eqs. (%
\ref{g12}) and (\ref{eff})]. In this case, algebraic equations (\ref{eta})
admit a simple solution, making it possible to eliminate the widths in favor
of the wave functions:
\begin{equation}
\eta _{1}^{4}=\eta _{2}^{4}=1+g(|f_{1}|^{2}+|f_{2}|^{2}),  \label{etaeta}
\end{equation}%
hence equations (\ref{f}) for the two coupled axial wave functions may be
cast into a closed form,
\begin{gather}
i\,\partial _{t}f_{k}=\Big[-\frac{1}{2}\partial
_{x}^{2}+V(x)+(-1)^{k-1}i\gamma \partial _{x}  \notag \\
+{\frac{1+{\left( 3/2\right) }g(|f_{1}|^{2}+|f_{2}|^{2})}{\sqrt{%
1+g(|f_{1}|^{2}+|f_{2}|^{2})}}}\Big]f_{k}+\Gamma \ f_{3-k}~.  
\label{ff}
\end{gather}%
This NPSE system is a generalization of the one introduced earlier \cite%
{sala-boris} for the study of vectorial solitons in two-component BECs,
under the same assumption (\ref{ggg}) as adopted here (but without the
linear coupling between the components).

\begin{figure}[t]
\begin{center}
{\includegraphics[width=8.5cm,clip]{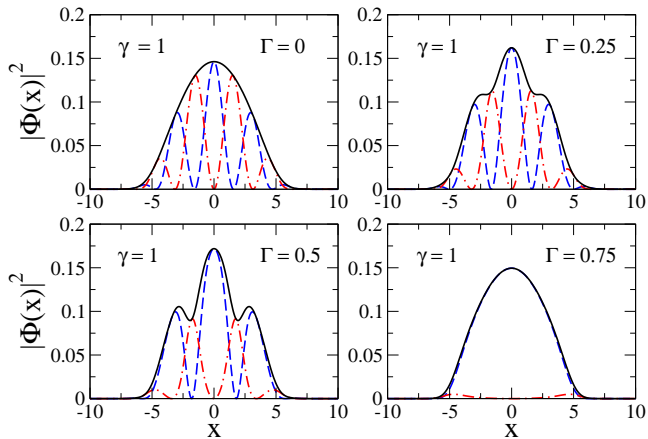}}
\end{center}
\caption{(Color online). The repulsive binary BEC under axial harmonic
confinement (\protect\ref{V}): The probability density for the trap
anisotropy $\protect\lambda =1/3$, adimensional nonlinearity strength $gN=20$%
, spin-orbit coupling $\protect\gamma =1$, and four values of the
adimensional Rabi coupling $\Gamma $. Here, and in similar plots displayed
below, the solid line depicts $|\Phi (x)|^{2}=\Phi _{R}^{2}(x)+\Phi
_{I}^{2}(x)$, while the dashed and dashed-dotted lines separately represent
squared real and imaginary parts of the wave function, $\Phi _{R}^{2}(x)\ $%
and $\Phi _{I}^{2}(x)$. Lengths are measured, here and below, in units of
the transverse confinement radius, $a_{\bot }=\protect\sqrt{\hbar /(m\protect%
\omega _{\bot })}$.}
\label{fig1}
\end{figure}

Using Eq. (\ref{ff}), we aim to construct stationary states with chemical
potential $\mu $, setting
\begin{equation}
f_{k}(x,t)=\phi _{k}(x)\ e^{-i\mu t}.\;  \label{phi}
\end{equation}%
The resulting equations for stationary fields $\phi _{1,2}(x)$ are
compatible with restriction
\begin{equation}
\phi _{1}^{\ast }(x)=\phi_{2}(x),  
\label{reduction}
\end{equation}%
leading to the single stationary NPSE:
\begin{equation}
\mu \Phi =-\frac{1}{2}\Phi ^{\prime \prime }+V(x)\Phi +i\gamma \Phi ^{\prime
}+{\frac{1+{(3/2)}gN|\Phi |^{2}}{\sqrt{1+gN|\Phi |^{2}}}\Phi }+\Gamma \Phi
^{\ast }\;,  
\label{figata}
\end{equation}%
where we have set $\Phi (x)\equiv \sqrt{2/N}\phi _{1}(x)=\sqrt{2/N}\phi
_{2}^{\ast }(x),$ so that $\int_{-\infty }^{+\infty }dx\ |\Phi (x)|^{2}=1$,
and the prime stands for $d/dx$. Clearly, solutions of Eq. (\ref{figata})
are complex if $\gamma \neq 0$, with the symmetry which implies that their
real and imaginary parts, $\Phi _{R}(x)$ and $\Phi _{I}(x)$, are,
respectively, even and odd functions, provided that potential $V(x)$ is even
(or is absent).

In fact, being interested in stationary solutions subject to constraint (\ref%
{reduction}), we can also consider a case more general than the one singled
out by Eq. (\ref{ggg}) (the full symmetry of the nonlinear interactions).
This generalization is essential because the case of the full symmetry is
critical, in some respects \cite{gg}. Assuming $\left\vert f_{1}\right\vert
^{2}=\left\vert f_{2}\right\vert ^{2}\equiv |f|^{2}$ and
\begin{equation}
g_{1}=g_{2}\equiv g\neq g_{12}~,  
\label{g12}
\end{equation}%
one can again obtain a simple solution for algebraic equations (\ref{eta}):
\begin{equation}
\eta _{1}^{4}=\eta _{2}^{4}=1+\left( g+g_{12}\right) |f|^{2},
\label{etaetaf}
\end{equation}%
the substitution of which into Eqs. (\ref{f}), (\ref{phi}) and (\ref%
{reduction}) leads to the stationary equation tantamount to Eq. (\ref{figata}%
), with $g$ replaced, according to Eqs. (\ref{g12}) and (\ref{etaetaf}), by
\begin{equation}
g_{\mathrm{eff}}\equiv (1/2)\left( g+g_{12}\right) .  
\label{eff}
\end{equation}
All the stationary solutions obtained 
from Eq. (\ref{figata}) pertain to the solutions 
subject to condition (\ref{reduction}), if $g$
is replaced by $g_{eff}$ in the case when relation (\ref{g12}) holds.
This equivalence does not apply to asymmetric solutions, which do not
obey relation (\ref{reduction}). Strictly speaking, 
the equivalence does not pertain 
either to the study of the dynamical solutions of the solutions, which must
be carried out withing the framework of full equations (\ref{ff}), and, in
particular, should include perturbations which 
may break relation (\ref{reduction}). 

Finally, we note that, in the weakly nonlinear regime, $gN|\Phi |^{2}\ll 1$,
the nonpolynomial term in Eq. (\ref{figata}) may be expanded and reduced to
the cubic approximation:
\begin{equation}
\left( \mu -1\right) \Phi =-\frac{1}{2}\Phi ^{\prime \prime }+V(x)+i\gamma
\Phi ^{\prime }+gN|\Phi |^{2}\Phi +\Gamma \Phi ^{\ast }.  \label{cubic}
\end{equation}
In the opposite limit of $gN|\Phi |^{2}\gg 1$, which may be relevant in the
case of the repulsive interactions, the nonpolynomial nonlinearity reduces
to a quadratic form, $\sim \left\vert \Phi \right\vert \Phi $, cf. Ref. \cite%
{Delgado}.

\subsection{Analytical approximations}

In the absence of the Rabi coupling, $\Gamma =0$, the SO term can be removed
from Eq. (\ref{figata}) by substitution
\begin{equation}
\Phi (x)\equiv \Phi _{0}(x) \, e^{i\gamma x},  \label{tilde}
\end{equation}%
the resulting equation for $\Phi _{0}(x)$ being tantamount to the stationary
version of the usual NPSE \cite{NPSE}, with a shifted chemical potential,
\begin{equation}
\tilde{\mu}\equiv \mu +\gamma ^{2}/2.  \label{shift}
\end{equation}%
The same substitution (\ref{tilde}) may be used to produce an analytical
result valid for small $\Gamma $ and/or large $\gamma $: in the lowest
approximation, the solution is%
\begin{equation}
\Phi (x)\approx \Phi _{0}(x)e^{i\gamma x}\left[ 1+\frac{i\Gamma }{2\gamma
^{2}}\sin \left( 2\gamma x\right) \right] ,  \label{approx0}
\end{equation}%
where $\Phi _{0}(x)$ is, as said above, a real solution for the usual NPSE
with $\Gamma =\gamma =0$ and chemical potential (\ref{shift}). Note that
this approximation predicts an increase of the height of the density profile
in the mode, averaged over oscillations between the real and imaginary parts
of the wave function:%
\begin{equation}
\overline{\left\vert \Phi (x)\right\vert ^{2}}\approx \Phi _{0}^{2}(x)\left(
1+\Gamma ^{2}/8\gamma ^{4}\right) .  \label{height}
\end{equation}

In the opposite case, when $\gamma $ is small and $\Gamma $ is large, an
analytical approach can be developed too. In this situation, a
straightforward consideration of Eq. (\ref{figata}) demonstrates that, in
the lowest approximation, the solution can be constructed as one with a
small imaginary part:%
\begin{equation}
\Phi (x)\approx \Phi _{0}(x)+\frac{i\gamma }{2\Gamma }\Phi _{0}^{\prime }(x),
\label{approx}
\end{equation}%
where $\Phi _{0}(x)$ is, as above, the solution of the usual NPSE
corresponding to the given norm, $N$.

Finally, a specific approximation applies to the description of broad
solitons, for which the kinetic-energy term in Eq. (\ref{ff}), $\left(
1/2\right) \partial _{x}^{2}f_{k}$, may be neglected in comparison with the
SO coupling, $\left( -1\right) ^{k-1}i\gamma \partial _{x}f_{k}$. Broad
solitons may be naturally assumed to have a small amplitude, hence the
accordingly expanded system of equations (\ref{ff}) is approximated by%
\begin{gather}
i\,\partial _{t}F_{k}=\Big[V(x)+(-1)^{k-1}i\gamma \partial _{x}  \notag \\
+{g}\left( \left\vert F_{1}\right\vert ^{2}+\left\vert F_{2}\right\vert
^{2}\right) \Big]F_{k}+\Gamma \ f_{3-k}~,  \label{F}
\end{gather}%
where $F_{k}\equiv f_{k}\exp \left( it\right) $. In the absence of the axial
potential ($V=0$), the self-attractive nonlinearity ($g<0$) gives rise to
well-known exact solutions of Eq. (\ref{F}) in the form of \textit{gap
solitons}, which may be moving ones, with velocity $c$ \cite%
{solitons,Martijn}, namely
\begin{eqnarray}
F_{1} &=&\frac{1}{\gamma }\sqrt{\Gamma \frac{\gamma -c}{2|g|}}\left( \gamma
^{2}-c^{2}\right) ^{1/4}\,W^{\ast }(X)\cdot   \notag \\
&&\mathrm{exp}\,\left[ i\phi (X)-iT\cos \theta \right] ,  \notag \\
F_{2} &=&\frac{1}{\gamma }\sqrt{\Gamma \frac{\gamma +c}{2|g|}}\left( \gamma
^{2}-c^{2}\right) ^{1/4}\,W(X)\cdot   \notag \\
&&\mathrm{exp}\,\left[ i\phi (X)-iT\cos \theta \right] ,  \label{GS}
\end{eqnarray}%
where
\begin{eqnarray}
X &=&\Gamma \left( \gamma ^{2}-c^{2}\right) ^{-1/2}\left( x-ct\right) ,\,
\notag \\
T &=&\left( \gamma ^{2}-c^{2}\right) ^{-1/2}\left( \Gamma /\gamma \right)
\left( \gamma ^{2}t-cx\right) ,  \notag \\
\phi (X) &=&\left( 2c/\gamma \right) ~\mathrm{\tan }^{-1}\left\{ \tanh \left[
(\sin \,\theta )X\right] \tan \left( \theta /2\right) \right\} ,  \notag \\
W(X) &=&\left( \sin \,\theta \right) \,\mathrm{sech}\left[ (\sin \,\theta
)X-i\left( \theta /2\right) \right] \,,  \label{W}
\end{eqnarray}%
and the parameter $\theta $ is determined by the normalization condition, $%
N=N_{1}+N_{2}$ [see Eq.(\ref{N})],
\begin{equation}
\theta =\left( |g|\gamma /2\right) \left( \gamma ^{2}-c^{2}\right) ^{-1}N.
\label{theta}
\end{equation}%
The solitons exist with velocities $c^{2}<\gamma ^{2}$, and with $\theta
<\pi $, which imposes a limitation on the number of atoms in the soliton of
this type, according to Eq. (\ref{theta}). In fact, the solitons are stable,
approximately, in the interval of $\theta <\pi /2$ \cite{stability}, which
makes the limitation on $N$ more restrictive.

The broad gaps solitons can be also found in the presence of the axial
trapping potential, $V(x)$, in Eq. (\ref{F}). Indeed, for the broad soliton
it may be approximated by $V(x)\approx -V_{0}\delta (x)$ with $V_{0}>0$.
Quiescent gap solitons (with $c=0$) can be found in an exact analytical form
in this case too \cite{Mak}.

The gap solitons exist as well in the case of the self-repulsion, i.e., $g>0$%
. The respective solutions are obtained from Eqs. (\ref{GS}) and (\ref{W}),
substituting $\left\{ F_{1},F_{2}\right\} $ by $\left\{ F_{1}^{\ast
},-F_{2}^{\ast }\right\} $.

Lastly, the underlying condition for neglecting the second derivatives in
Eq. (\ref{ff}) and reducing it to Eq. (\ref{F}) takes the form of $N\ll
\left( \Gamma |g|\right) ^{-1}\left( \gamma ^{2}-c^{2}\right) ^{3/2}$. Of
course, the gap solitons do not exist in the rigorous sense, as the
neglected second derivatives close the spectral gap, giving rise to a slow
decay of the solitons into radiation waves, cf. Ref. \cite{Alan}.

\section{Numerical results}

\subsection{The self-repulsive spin-orbit-coupled BEC under the
harmonic-oscillator axial confinement}

\begin{figure}[t]
\begin{center}
{\includegraphics[width=8.5cm,clip]{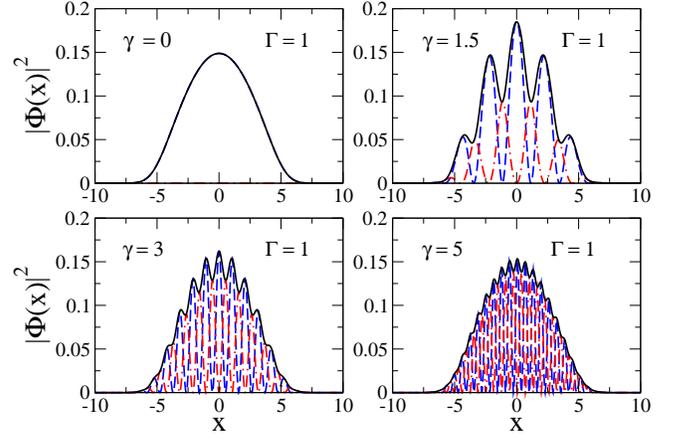}}
\end{center}
\caption{(Color online). The same as in Fig. \protect\ref{fig1}, but for a
fixed Rabi coupling, $\Gamma =1$, and four different values of the
spin-orbit coupling, $\protect\gamma $.}
\label{fig2}
\end{figure}

We start the numerical analysis by looking for trapped modes supported by
Eq. (\ref{figata}) with $g>0$ (the repulsive nonlinearity) in the presence
of the HO axial potential with frequency $\omega _{x}$,
\begin{equation}
V(x)={(\lambda ^{2}/2)}x^{2},  \label{V}
\end{equation}%
where $\lambda \equiv \omega _{x}/\omega _{\bot }$ is the anisotropy of the
HO confinement. Equation (\ref{figata}) was solved by dint of the
imaginary-time method, implemented with the help of the finite-difference
predictor-corrector Crank-Nicolson algorithm \cite{sala-numerics}. In Fig. %
\ref{fig1} the numerical results are reported for the HO potential (\ref{V})
with $\lambda =1/3$, adimensional nonlinearity strength $gN=20$, SO coupling
$\gamma =1$, and four different values of Rabi coupling $\Gamma $. The
density profile of the trapped modes, $|\Phi (x)|^{2}$, is plotted as a
function of axial coordinate $x$ (solid lines). For the completeness of the
depiction of the complex wave functions, we also display $\Phi _{R}^{2}(x)$
and $\Phi _{I}^{2}(x)\ $(dashed and dot-dashed lines, respectively). As
expected, for $\Gamma =0$ (the left upper panel of Fig. \ref{fig1}) density $%
|\Phi (x)|^{2}$ for $\gamma =1$ is the same as for $\gamma =0$. In this
case, a finite value of $\gamma $ implies that $\Phi (x)$ has oscillatory
real and imaginary components $\Phi _{R}(x)$ and $\Phi _{I}(x)$. The
increase of $\Gamma $ at fixed $\gamma $ ($\gamma =1$ in Fig. \ref{fig1})
leads to the decrease of the imaginary part, the corresponding density
profile, $|\Phi (x)|^{2}$, displaying several local maxima (see the right
top and left bottom panels in Fig. \ref{fig1}). Finally, when $\Gamma $ is
sufficiently large, the imaginary component of the density, $\Phi _{I}^{2}(x)
$, becomes very small, in accordance with analytical approximation (\ref%
{approx}), the probability density $|\Phi (x)|^{2}$ being nearly identical
to that at $\Gamma =0$, as can be seen from the comparison of the right
bottom and left top panels in Fig. \ref{fig1}).

It is also relevant to analyze the density profile at fixed $\Gamma $ for
different values of the SO coupling $\gamma $. To this end, in Fig. \ref%
{fig2} we display the numerical results for $\lambda =1/3$, $gN=20$, $\Gamma
=1$, and four values of $\gamma $. At $\gamma =0$ (the left upper panel of
Fig. \ref{fig2}), $\Phi (x)$ is real and the density profile is smooth. As
shown in Fig. \ref{fig2}, for $\gamma \neq 0$ solution $\Phi (x)$ is
complex, and the respective density $|\Phi (x)|^{2}$ displays several local
maxima, whose number increases with $\gamma $. At large values of $\gamma $,
we observe many small-amplitude local variations of $|\Phi (x)|^{2}$, which
makes the averaged density profile close to that at $\gamma =0$. This regime
may be explained by analytical approximation (\ref{approx0}).

\begin{figure}[t]
\begin{center}
{\includegraphics[width=8.5cm,clip]{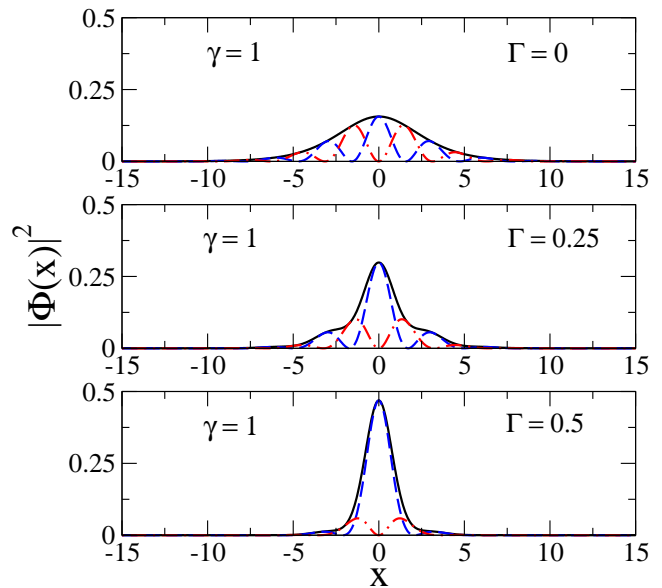}}
\end{center}
\caption{(Color online). Density profiles of the solitons, in the case of
the self-attractive binary BEC without the axial potential, $V=0$. Here, the
nonlinearity strength is $gN=-0.6$, spin-orbit coupling is $\protect\gamma =1
$, and values of the Rabi coupling, $\Gamma $, are indicated in the panels.}
\label{fig3}
\end{figure}

\begin{figure}[t]
\begin{center}
{\includegraphics[width=8.cm,clip]{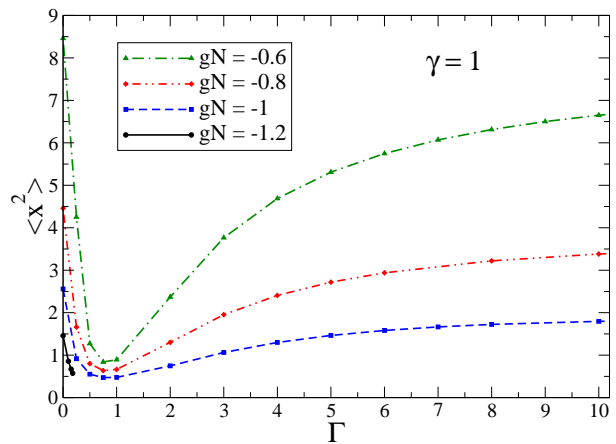}}
\end{center}
\caption{(Color online). The axial squared width $\langle x^{2}\rangle $ of
the solitons, defined as per Eq. (\protect\ref{<>}), versus the Rabi
coupling, $\Gamma $. Here the spin-orbit coupling is $\protect\gamma =1$,
and four values of the nonlinearity strength, $gN$, are indicated in the box
(the curve for $gN=-1.2$ is very short, terminating because of the onset of
the collapse).}
\label{fig4}
\end{figure}

To summarize, the numerical results demonstrate that the SO and Rabi
couplings produce the most prominent effect on the density profile of the
trapped modes in the self-repulsive binary BEC, in the form od conspicuous
side lobes, in the case when the constant accounting for both couplings, $%
\gamma $ and $\Gamma $, take moderate values, neither very small nor too
large. It is also relevant to stress that Figs. \ref{fig1} and \ref{fig2}
demonstrate values $gN\left\vert \Phi (x=0)\right\vert ^{2}\simeq 3$, at
peak-density points. This implies, according to Eq. (\ref{figata}), that the
nonpolynomiality of the effective nonlinearity is quite essential.

\subsection{Bright solitons in the self-attractive spin-orbit-coupled BEC}

It is well known that the attractive intrinsic nonlinearity ($g<0$) supports
self-trapped matter-wave solitons in the effectively 1D BEC, even in the
absence of the axial confinement [$V(x)=0$] \cite{soliton}. Here we aim to
investigate effects of the SO and Rabi coupling on the bright solitons by
solving Eqs. (\ref{ff}) and (\ref{figata}) with $g<0$, in the absence of the
axial trapping potential [$V(x)=0$]. The objective is to construct true
solitons, unlike the approximate gap solitons (\ref{GS}).

For $\gamma =0$ and/or $\Gamma =0$, Eqs. (\ref{ff}) and (\ref{figata}) with $%
V(x)=0$ predict that metastable solitons exist only for $-{4/3}<gN<0$.
Indeed, at $gN<-4/3$ the 1D NPSE\ gives rise to the collapse of the
condensate, which sets in locally at a point where the total density of the
condensate attains the critical value, $\left\vert f_{1}\right\vert
^{2}+\left\vert f_{2}\right\vert ^{2}=-1/g$ [or $\left\vert \Phi \right\vert
^{2}=-\left( gN\right) ^{-1}$, in terms of Eq. (\ref{figata})]. This
prediction is known to be in very good agreement with the full 3D
calculations \cite{NPSE,NJP}.

Obviously interesting issues are a change of the shape of the solitons, and
a shift of the collapse threshold under the action of the SO and Rabi
couplings. In Fig. \ref{fig3} we plot the density profile $|\Phi (x)|^{2}$
(solid line) for the bright solitons, obtained from the numerical solution
of Eq. (\ref{figata}), for $gN=-0.6$, $\gamma =1$, and three values of $%
\Gamma $. A noteworthy feature is the compression of the soliton's density
profile and increase of its height with the increase of $\Gamma $. This
trend can be explained by the above-mentioned perturbative result, given by
Eq. (\ref{height}).

At a certain finite value of $\Gamma $, the bright soliton reaches its
smallest axial width and largest peak density, $\left\vert \Phi \left(
x=0\right) \right\vert ^{2}$, while the imaginary component (the
dashed-dotted line) of $\Phi (x)$ becomes small, see the lower panel of Fig. %
\ref{fig3}. With the further increase of $\Gamma $, the imaginary part of
the wave function vanishes, in accordance with Eq. (\ref{approx}), while the
axial width approaches a finite asymptotic value, corresponding to the real
solution with given norm $N$. This trend is displayed in Fig. \ref{fig4},
which shows the average soliton's width,%
\begin{equation}
\left\langle x^{2}\right\rangle \equiv N^{-1}\int_{-\infty }^{+\infty
}\left\vert \Phi (x)\right\vert ^{2}dx,  \label{<>}
\end{equation}%
versus $\Gamma $. The four curves with symbols of Fig. \ref{fig4} correspond
to different values of the adimensional nonlinearity strength $gN$. Note
that for $gN=-1.2$ (the solid line with filled circles) the collapse of the
binary BEC happens at $\Gamma \approx 0.19$, therefore this line is aborted
in Fig. \ref{fig4}. Thus, at fixed values of the SO coupling ($\gamma =1$ in
Fig. \ref{fig4}) a finite Raman coupling $\Gamma $ corresponds to the lowest
collapse threshold.

\begin{figure}[t]
{\LARGE {$|g|N$} }
\par
\begin{center}
{\LARGE {\includegraphics[width=8.cm,clip]{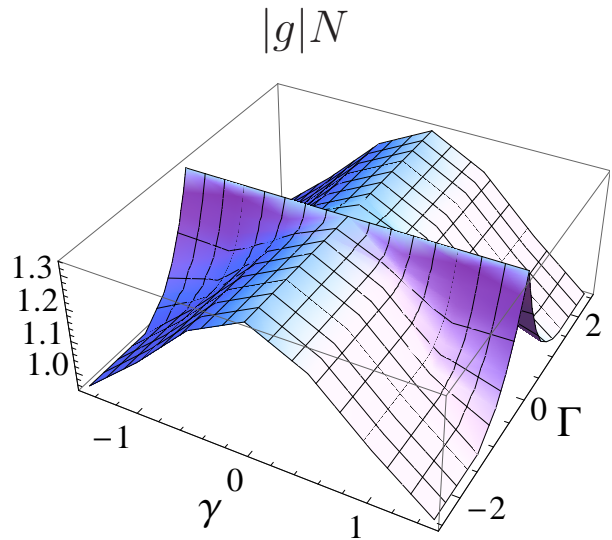}} }
\end{center}
\caption{(Color online). The critical value of the nonlinearity strength,
corresponding to the onset of the collapse in the 1D soliton, versus the
adimensional spin-orbit and Rabi couplings, $\protect\gamma $ and $\Gamma $.
}
\label{fig5}
\end{figure}

The results are summarized in Fig. \ref{fig5}, which shows the collapse
threshold in the parameter space $(\gamma ,\Gamma ,|g|N)$ for the
self-attractive binary BEC in the absence of the axial confinement. The
collapse points are obtained from numerical solutions of Eq. (\ref{figata}),
as those beyond which no solution can be found (we here do not study
dynamics of the collapsing states). We stress that, in accordance with Fig. %
\ref{fig4}, the collapse is first attained at a \textit{finite value} of $%
\Gamma $, close to the point of the minimum of the soliton's width, rather
than at $\Gamma \rightarrow \infty $. Furthermore, Fig. \ref{fig5} clearly
shows that the SO or Rabi coupling, acting in isolation, do not affect the
collapse threshold at all, while the strongest reduction of the collapse
threshold is produced by the interplay of these two interactions added to
the binary condensate (i.e., in terms of Fig. \ref{fig5}, the threshold
features the steepest descent, roughly, in the diagonal direction).

To illustrate the symmetry of the system, the figure includes both positive
and negative values of $\gamma $ and $\Gamma $. Note that negative $\Gamma $
corresponds to the fact that, instead of Eq. (\ref{reduction}), one could
impose constraint $\phi _{1}^{\ast }(x)=-\phi _{2}(x)$, which gives Eq. (\ref%
{figata}) with $\Gamma $ replaced by $-\Gamma $. While, as stressed above,
for $\Gamma >0$ the real and imaginary parts of the soliton's wave function,
$\Phi _{R}(x)$ and $\Phi _{I}(x)$, are even and odd, respectively, they have
opposite parities for $\Gamma <0$. Further, the critical surface in Fig. \ref%
{fig5} is symmetric with respect to the change of $\gamma \rightarrow
-\gamma $ because this is tantamount to $x\rightarrow -x$.

\section{Conclusion}

We have derived the 1D system of coupled Gross-Pitaevskii equations
for the relatively dense binary BEC, in which the two components are
coupled by the SO (spin-orbit) and Rabi linear terms, and also by
the nonlinear interactions, both self0repulsive and self-attractive.
The analysis was focused on effects caused by the nonpolynomial
character of the nonlinearity in the 1D equations derived from the
system of 3D GPEs. In addition to the numerical results, analytical
approximations were developed too, for the cases when the SO and
Raman couplings are weak, as well as for broad trapped modes. The
results demonstrate essential changes of the shape of the modes due
to the nonpolynomial nonlinearity, which can be partly explained by
means of the analytical approximations. In the case of the
self-repulsive nonlinearity, the interplay of the SO and Rabi
couplings cause a strong deformation of the shape of the trapped
modes, which develop large side lobes. The most significant result
is the reduction of the strength of the self-attractive nonlinearity
at the collapse threshold for the effectively 1D solitons under the
combined action of the SO and Rabi couplings.

An interesting possibility is to extend the present analysis for effectively
2D binary BEC in the presence of a tight 1D trapping potential acting in the
transverse direction. In the absence of the linear couplings, 2D
single-component solitons affected by the corresponding nonpolynomial
nonlinearity were studied in Ref. \cite{we-2D}. Here too, one may expect a
nontrivial lowering of the collapse threshold for the effectively 2D
solitons.

\section*{Acknowledgments}

LS thanks for partial support Universit\`a di Padova (Research Project
"Quantum Information with Ultracold Atoms in Optical Lattices"), Cariparo
Foundation (Excellence Project "Macroscopic Quantum Properties of Ultracold
Atoms under Optical Confinement"), and Ministero Istruzione Universita
Ricerca (PRIN Project "Collective Quantum Phenomena: from
Strongly-Correlated Systems to Quantum Simulators").

\end{document}